\documentstyle[emulateapj]{article}
\def\ie{i.e.\ }
\def\eg{e.g.,\ }
\def\etal{et~al.\ }
\def\ltsima{$\; \buildrel < \over \sim \;$}
\def\simlt{\lower.5ex\hbox{\ltsima}}
\def\gtsima{$\; \buildrel > \over \sim \;$}
\def\simgt{\lower.5ex\hbox{\gtsima}}

\def\h1{\ion{H}{1}\ }

\def\h2{H$_2$}
\def\coh2{CO/H$_2$}
\journalid{}{}
\articleid{}{}
\lefthead{Holley-Bockelmann \etal}
\righthead{The Evolution of Cuspy Triaxial Galaxies Harboring Central Black Holes}
 
\begin{document}
 
\title{The Evolution of Cuspy Triaxial Galaxies Harboring Central Black Holes}
 
\author{Kelly Holley-Bockelmann,\altaffilmark{1} J. Christopher Mihos,\altaffilmark{2} 
Steinn Sigurdsson,\altaffilmark{3}, Lars Hernquist\altaffilmark{4}, and Colin Norman \altaffilmark{5}}

\altaffiltext{1}{Department of Astronomy, University of Massachusetts,
Amherst, MA, 01002, kelly@shrike.astro.umass.edu}
\altaffiltext{2}{Department of Astronomy, Case Western Reserve University,
10900 Euclid Ave, Cleveland, OH 44106, hos@burro.astr.cwru.edu}
\altaffiltext{3}{Department of Astronomy and Astrophysics,
Penn State University, 525 Davey Lab,
University Park, PA 16802, steinn@astro.psu.edu}
\altaffiltext{4}{Department of Astronomy, Harvard University,
60 Garden St, Cambridge, MA 02138, lhernqui@cfa.harvard.edu}
\altaffiltext{5}{Department of Physics and Astronomy, Johns Hopkins University,
Baltimore, MD 02138, norman@stsci.edu}
 
\begin{abstract}

We use numerical simulations to study the evolution of triaxial
elliptical galaxies with central black holes.  In contrast to earlier
studies which used galaxy models with central density ``cores,'' our
galaxies have steep central cusps, as observed in real ellipticals.
As a black hole grows in these cuspy triaxial galaxies, the inner
regions become rounder owing to chaos induced in the orbital families
which populate the model.  At larger radii, however, the models
maintain their triaxiality, and orbital analyses show that
centrophilic orbits there resist stochasticity over many dynamical
times.  While black hole induced evolution is strong in the inner
regions of these galaxies, and reaches out beyond the nominal ``sphere
of influence'' of a black hole, our simulations do not show evidence
for a rapid {\it global} transformation of the host.  The triaxiality
of observed elliptical galaxies is therefore not inconsistent with the
presence of supermassive black holes at their centers.

\end{abstract}
 
\keywords{galaxies: elliptical, galaxies: kinematics and dynamics, 
galaxies: structure, methods: N-body simulations}
 
\section{Introduction}

Observations indicate that most, and perhaps all elliptical galaxies
harbor supermassive black holes at their centers (e.g. Gebhardt et al.
2000; Richstone et al. 1998; but see Gebhardt et al. 2001b).  In fact,
best-fit models of black hole demography indicate that roughly $97\%$
of ellipticals contain such black holes (Magorrian et al. 1998).  The
masses of these black holes seem to be correlated with properties of
the host bulge; current dynamical estimates have yielded black hole
masses of order $0.005 M_{\rm bulge}$ (Kormendy \& Richstone 1995;
Magorrian et al. 1998; van der Marel 1999).  There is also apparently
a trend between black hole mass and galaxy velocity dispersion,
implying that there is a Fundamental Plane even in the
four-dimensional space defined by [$ \log M_{\rm BH}, \log L, \log
\sigma_e, \log R_e$] (Gebhardt et al.  2001a; Ferrarese \& Merritt
2001).  These correlations suggest that galaxy formation and the
formation of central black holes are deeply connected.

It is believed that a central supermassive black hole can profoundly influence the evolution
of its host galaxy.  Massive black holes dominate the
galactic potential inside a radius of influence $ r_{\rm BH} \approx 100
(M_{{\rm BH}, 9}/ \sigma_{0,200}^2)$ pc where $M_{{\rm BH},9}$ is the
black hole mass in units of $10^9 M_\odot$ and $\sigma_{0,200}$ is the
central velocity dispersion in units of $200$ km/sec. Within
$r<r_{\rm BH}$, the three-dimensional structure and phase space of a
galaxy will be determined largely by the black hole.  Moreover, the
effects of a central black hole can reach far beyond $r_{\rm BH}$.  Owing to
discrete encounters with individual stars, central black holes will
``wander,'' in a manner akin to molecular Brownian motion (see,
e.g. Chatterjee, Hernquist \& Loeb [2001] for a recent analysis of
this phenomenon), effectively increasing $r_{\rm BH}$.  During galaxy
cannibalism, supermassive black holes disrupt even dense low-mass
companions in radial encounters (Holley-Bockelmann \& Richstone 2000;
Merritt \& Cruz 2001) and may tidally torque debris into a
nuclear disk following mergers from eccentric orbits
(Holley-Bockelmann \& Richstone 2000).  Black hole binaries created
during galaxy mergers may explain the flat density profiles seen in
the inner regions of the largest ellipticals (Makino \& Ebisuzaki
1996; Quinlan \& Hernquist 1997).

Central black holes can induce secular evolution as well.  According
to one long-standing suggestion, the growth of a massive black hole in
a triaxial potential can destabilize centrophilic box orbits through
stochastic diffusion, driving the global shape of a galaxy towards
axisymmetry in a few crossing times (Gerhard \& Binney 1985; Norman,
May, \& van Albada 1985; Merritt \& Quinlan 1998; Wachlin \&
Ferraz-Mello 1998; Valluri \& Merritt 1998).  If this picture of black
hole induced evolution away from triaxiality is correct, it will have
serious ramifications for our understanding of the fueling of active
galactic nuclei (AGN) and the observed correlations between the
properties of black holes and their host galaxies.  Indeed, such
evolution could lead to a self-regulation of AGN activity and black
hole growth in ellipticals.  If triaxiality is linked to AGN fueling,
as a consequence of gas being unable to settle into closed orbits in
such a potential (Norman \& Silk 1983), then an evolution towards
axisymmetry would naturally lead to a reduction in the feeding of a
central black hole.  Such a scenario might also explain, at least in
part, the correlation between black hole and bulge mass in early type
galaxies (Valluri \& Merritt 1998).

While theoretical arguments paint a compelling picture for a link
between central black holes and the loss of triaxiality in galaxies,
observational support remains somewhat problematic, 
owing to the difficulty of
inferring the three dimensional structure of galaxies from their
projected properties.  However, recent studies indicate that at least
some luminous ellipticals are triaxial (Binney 1976; Franx \etal 1991;
Tremblay \& Merritt 1995; Bak \& Statler 2000).  The triaxiality of 
these systems indicates either
that they contain at most low mass black holes (insufficient to drive
evolution) or that the coupling between black holes and galaxy shape
is not well understood.  The former possibility appears to be
inconsistent with the notion that luminous AGN are preferentially
found in luminous ellipticals.

These considerations emphasize the need for realistic models of
galaxies containing central black holes.  Earlier theoretical studies
have employed either triaxial models with constant density cores
(Norman et al. 1985; Merritt \& Quinlan 1998) or spherical models with
central density cusps (e.g. Sigurdsson et al. 1995).  Neither of these
two choices is entirely satisfactory.  Observations of ellipticals
made with the Hubble Space Telescope show that these galaxies
typically have central density cusps $\propto r^{-\gamma}$ with
$\gamma=0.5-2$ (Lauer et al. 1995).  Galaxies with density cusps
support different stellar orbits than e.g.  $\gamma=0$ core models
(Gerheard \& Binney 1985; Gerhard 1986; Pfenniger \& de Zeeuw 1989; Schwarzschild 1993; de Zeeuw 1996; Merritt 1999; Holley-Bockelmann et al. 2001), 
altering the response of a galaxy to
the growth of a black hole.  Likewise, spherical galaxies will support
different orbital families than triaxial ones, whether or not they
have central density cusps.  While simulations employing spherical
models have demonstrated the growth of black hole induced cusps and
the polarization of the velocity ellipsoid in such objects (Sigurdsson
et al. 1995), they of course have little to say about the evolution of
{\it triaxial} ellipticals.

To address these shortcomings, we present a study of black hole growth
in triaxial elliptical galaxies with central density cusps, using the
``adiabatic squeezing'' technique described in Holley-Bockelmann et
al.  (2001; hereafter, ``Paper 1'') to generate models with prescribed
shapes and cusps that are stable for many dynamical times.  Using
N-body simulations, we then adiabatically grow a black hole in these
galaxies over several crossing times.  We characterize the orbital
families which populate the models, and find that many of the highly
bound boxes, boxlets, and eccentric tubes are transformed into chaotic
orbits -- a clear signature of the influence of the spherical central
potential in a larger, nearly unchanged, triaxial figure.

The paper is organized as follows.  Section 2 summarizes our technique
for generating a stable triaxial model of a galaxy, with subsequent
growth of a central black hole.  Section 3 shows the outcome of black
hole growth in a model with central density cusp $\gamma=1.0$ and
initial half mass axis ratios $b/a, c/a = 0.85, 0.7$, and 
discusses the structural changes occurring in this model as a result of
the growth of a black hole of mass $M_{\rm BH}= 0.01 M_{\rm gal}$.
Section 4 explores the evolution of the orbital population induced by
the black hole, and Section 5 provides conclusions.

\section{Modeling Technique}

The ``adiabatic squeezing'' technique for generating triaxial models
is discussed in detail in Paper 1.  To summarize, we begin with,
e.g. a spherical Hernquist (1990) model, having a density profile
$$\rho(r)={M\over 2\pi}{a\over r}{1\over (r+a)^3},$$ where $M$ is the
total mass and $a$ is a scale length.  We generate an N-body
realization of this spherical model using the multimass scheme of
Sigurdsson et al.  (1995), so that particles have a mass which is
roughly inversely proportional to their pericentric radius.  This
technique gives a finer sampling of the phase space of tightly bound
orbits than would be feasible if equal mass particles were used
throughout.  We then evolve this model while adiabatically applying a
drag force first along the $z$ axis (using the SCF method described
below to maintain axisymmetry in the $xy$ plane) and then along the
$y$ axis to yield a triaxial figure.  The system is rescaled after
each step so that the scale radius of the long axis, $a$, is
unity. The lengths of the other axes are correlated with the strength
of the drag force applied in each direction.  To ensure an equilibrium
final state, the model is evolved for several half-mass dynamical
times without any drag forces present.

A black hole of final mass $M_{\rm BH}$ is then grown adiabatically in one
of these equilibrium models over a time $t_{\rm BH}$ according to:
$$ M(t) = \cases{{M_{\rm BH}} \Bigl[ 3 {{\bigl( {{t} \over {t_{\rm BH}}} \bigr)}^2} - 2 \bigl( {{{t} \over {t_{\rm BH}}} \bigr)} ^3 \Bigr]  ,&for $t \le t_{\rm BH}$;\cr
                 M_{\rm BH}, & for $t > t_{\rm BH}$.\cr}$$
This expression ensures smooth evolution of the black hole mass,
\ie that $\dot M(t) = 0$ at  $t=0$ and at $t=t_{\rm BH}$.
The potential of the black hole is that of a softened point mass:

$$\Phi_{\rm BH}(r,t) = {M(t) \over \sqrt { r^2 + \epsilon_{\rm BH}^2 }},$$

\noindent where the softening parameter is set to
$\epsilon_{\rm BH}= 0.001$ in the simulations described here. This
softening sets the resolution scale of our model and can be compared
to the radius of influence of a 1\% mass black hole, which is
$r_{\rm BH} \sim 0.03$.

To ensure that black hole growth is adiabatic, we chose $t_{\rm BH}$
to be long compared to the orbital periods of stars in the inner
regions of the galaxy (Sigurdsson et al. 1995).  We varied the value
chosen for $t_{\rm BH}$ and determined that for the examples presented
here a black hole growth timescale $t_{\rm BH}=20$ is adequate.  This
can be compared with the dynamical times for the fiducial Hernquist
sphere with $a=1$ which are $3.14, 5.96,$ and $8.33$ at the scale
radius ($r=1.0$), effective radius ($r=1.815$), and half-mass radius
($r=2.414$), respectively.  After the black hole has reached its final
mass, the model is evolved for another 20 time units to allow the
model to reach equilibrium.

The simulations were performed with a self-consistent field (SCF)
code (e.g. Hernquist \& Ostriker 1992; Hernquist et al. 1995). In
this approach, the density and potential of the galaxy are expanded in
a set of basis functions, with the lowest order term chosen to
represent the underlying density profile.  The expansion coefficients
are determined from the particle distribution, using $n$ radial terms
and ($l,m$) angular terms.  The examples here adopted $n_{max}=10$ and
$m_{max}=l_{max}=6$.  The $N$-body particles move in the combined
potential field of the SCF expansion and the central black hole
(Sigurdsson et al. 1997), and orbital accuracy is ensured by using a
high-order Hermite integrator with variable time-stepping.  The
simulations were run on the T3E and the Blue Horizon at the San Diego
Supercomputer Center.

\section{Results of Dynamical Modeling}

We demonstrate our technique with an N=512,000 particle Hernquist
model with total mass $M=1$ and initial axis ratios $b/a=0.85$ and
$c/a=0.7$, measured at the half mass radius.  To this model, we add an
adiabatically growing black hole of final mass $M_{\rm BH}= 0.01$.  If
we insist that this model lies on both the global and core Fundamental
Planes (Faber et al. 1997), then the central density slope of
$\gamma=1$ (neglecting the influence of the black hole) fixes the
absolute magnitude ($M_{\rm V}\approx-21.6$), effective radius 
($r_{\rm eff} \approx 6$ kpc),
and projected central velocity dispersion ($\sigma_{p} \approx 310$ km/sec) of
the corresponding galaxy.  Scaling our model to such a galaxy results
in a unit length, unit velocity, and unit time corresponding to $3$
kpc, $1200$ km/sec, and $3\times10^6$ years, respectively. Thus, in
our model, the scaled half-mass dynamical timescale is $2.5 \times
10^7$ years.

As the black hole grows, both the cusp slope, $\gamma$, and projected central
velocity dispersion, $\sigma_p$, increase.  Figure 1 shows the cusp
slope (measured at ellipsoidal radius $\log q = -1.3$) and the
projected central velocity dispersion of the model (measured at
projected ellipsoidal radius $\log Q = -2.3$) as a function of time
during and after black hole growth.  After an initial rapid increase
in $\gamma$ during the black hole growth phase (between $0<T<20$), the
evolution in both $\gamma$ and $\sigma_p$ all but ceases, settling at
the equilibrium values $\gamma \simeq 2.05$, $\sigma_p \simeq 0.7$.
These results are characteristic of adiabatic black hole growth in
cuspy galaxies and can be compared both to analytic estimates for
adiabatic black hole growth in a spherical $\gamma=1.0$ model, which
predict $\gamma = 7/3$, $\sigma_p=0.75$ (Quinlan, Hernquist \&
Sigurdsson 1995) and to the results from N-body simulations of
adiabatic black hole growth in a Hernquist sphere, where $\gamma
\approx 2.2$ and $\sigma \approx 0.65$ (Sigurdsson, Hernquist, \&
Quinlan 1995).\footnote{The fact that our measured cusp slope is less
than the analytic value of $\gamma = 7/3$ is to be expected since our
cusp slope is measured over a finite radial range near the center, and
is not the asymptotic $r=0$ value.}  Note that these high central
velocity dispersions are measured on length scales smaller than the
typical minimum spatial resolution at the distance of nearby
ellipticals; observable central velocity dispersions are averaged over
larger apertures and tend to obscure velocity cusps. 

The evolution in galaxy shape is shown in Figure 2 for the innermost
2\%, 10\%, and 50\% of the galaxy (by mass).  Prior to black hole
growth, the model has a more or less uniform shape throughout, with a
slight tendency towards greater triaxiality in the center (see Figure
2 at $T=0$ and Paper 1).  As the black hole grows, the inner regions
quickly become rounder (Figure 2, panel a); in fact, by the time the
black hole has reached full mass, the central 10\% of the mass,
corresponding to an ellipsoidal radius $q$ $=\sqrt{ x^2 + (y/b)^2 +
(z/c)^2}< 0.1$ is practically spherical with axis ratios $a:b:c=
1.0:0.95:0.92$.  The shape evolution in the outer regions is much less
dramatic, however (Figure 2, panel c).  Following the growth of the
black hole, the model exhibits a marked shape gradient, becoming more
strongly triaxial with increasing radius.  This shape profile,
essentially axisymmetric in the center and a relatively unaltered
triaxial figure further out, remains stable as the system settles into
equilibrium from $20<T<40$, even though the dynamical timescale here
is much shorter than the integration time of the simulation. Despite the
near sphericity at the center, this region is, perhaps surprisingly,
 still triaxial enough to influence the stellar orbital dynamics  
(Statler 1987; Hunter \& de Zeeuw 1992), at least in principle.
We will return to this issue of orbits in the next section.

The final state of this model features several hallmarks of a black
hole-embedded triaxial figure.  Figure 3 shows the properties of this
object as a function of ellipsoidal radius $q$ at $T=40$ (12.8$t_{\rm
dyn}$ at $q=1$), well after the black hole has stopped growing.
Figure 3a shows the $\gamma\sim 2$ density cusp induced by the black
hole inside $\log q = -1$.  At a larger radii $\log q > -1$, though,
this plot demonstrates that the system retains the original Hernquist
density profile. Figure 3b shows explicitly the strong shape gradient
in the model.  Inside $r_{\rm BH}$, both the projected and intrinsic
velocity dispersions exhibit a strong central cusp (panels c and
d). In the outskirts, where the model maintains its triaxiality, the
projected velocity distributions follow $\sigma_x>\sigma_y>\sigma_z$ in
accord with a triaxial model where $a>b>c$. However, inside the cusp
the projected velocity dispersions are commensurate. Interestingly,
the anisotropy parameter, $\beta = 1-\langle v_t^2\rangle / \langle
v_r^2\rangle$, becomes negative near the black hole, as many radial
orbits are given a large tangential component. This is consistent with
models of stellar orbits around a black hole that is adiabatically
grown, where $\beta=-0.3$ (Goodman \& Binney 1983; Quinlan et
al. 1995). Exterior to the black hole's radius of influence, the system
is radially anisotropic ($\beta>0$), as expected for a triaxial galaxy.

Structurally, the most dramatic change induced in the model by the
black hole is the strong shape gradient as one moves inwards (from
$b,c= 0.9, 0.8$ at $\log q = 0$ to $b,c = 0.95,\ 0.92$ at $\log
q=-2$).  This significant central roundening may arise if box, boxlet, and
eccentric tube orbits are lost to stochasticity near the black hole (e.g. 
Norman et al. 1985; Gerhard \& Binney 1985), yet survive at larger radii.
However, even the nearly spherical central regions may host
significantly triaxial dynamics, as pointed out by Statler (1987) and
Hunter \& de Zeeuw (1992). Clearly, a more rigorous orbital analysis is 
warranted; we present this analysis in the next section.

\section{Orbital Properties}

In equilibrium, the structure of a galaxy and its orbital phase space
are closely related via the Collisionless Boltzmann equation.  
Since the intrinsic shape
of an elliptical is dictated by the time-averaged configuration space
density of the orbits of its stars, the shape of the galaxy may change
with time, or with radius, in response to a changing potential.  Hence
information can be gleaned about galactic structure through the
analysis of orbit families and their subsequent stability in our
model.  Both two- and three-dimensional analyses are useful: planar
orbits are subject to less numerical scattering and provide surfaces
of section that are easy to analyze and can be compared to previous
studies, while three-dimensional orbits directly trace the particle
distribution and populate resonances that do not project cleanly onto
a plane.

Using the automated orbit classifier described in Paper 1, we analyzed
both the two-dimensional phase space along the $xy$ and $xz$ planes
(where, as always, we take $x$, $y$, and $z$ to be aligned with the
major, intermediate, and minor axes respectively) and the
three-dimensional phase space populated by the particles at $T=40$.
We remind the reader that the initial state of this model (the
pre-black hole stage) is the same model analyzed in Paper 1, so a
direct determination of the effect of the black hole can be made by
contrasting the figures in Paper 1 to those here.  As before, we
exploited the 8-fold symmetry of the potential to minimize noise in
the particle distribution. Orbits were typed by the resonances in the
dominant Fourier frequencies of the particle's motion in each plane
(see Paper 1 for more information on these techniques). Following
previous work, we identified stochasticity by searching for a
significant change in the fundamental frequency over two time
intervals (Paper 1; Valluri \& Merritt 1998).  Each time interval was
comprised of 50 orbital periods of a long axis tube at the energy of
the orbit in question.\footnote{Hence, each integration interval
corresponds to 200 dynamical times. We note that Paper 1 erroneously
stated that each time interval corresponds to 12.5 dynamical times;
however, this misstatement did not affect any of the conclusions in
that paper.}  Each orbit was integrated for a total of $\approx 200$
orbital times, or until a hard limit of $T=8000$. This hard limit was
chosen to correspond to greater than a Hubble time of evolution in the
model when it is scaled to real galactic units for a galaxy at
$M_{\rm V}=-21.6$. At $3 R_e$ in such a galaxy, the total integration is
$\approx$ two Hubble times, where each integration interval is over 11
Gyr at this hard limit.

We have simplified our identification of strongly chaotic orbits
somewhat from the method used in Paper 1. Before, we followed the
convention of Valluri \& Merritt (1998) to define chaotic orbits.
Specifically, an orbit was considered strongly chaotic if $\Delta f =
\mid f_1 - f_2 \mid/ f_0 T > f_{\rm crit}$, where $f_1$ and $f_2$ are
the dominant frequencies at the first and second time intervals, $f_0$
is the frequency of a tube about the long axis, T is the time between
the sampled intervals, and $f_{\rm crit}$ is the critical threshold
for the onset of a strongly chaotic orbit. For the black hole-less
model, we set the following chaotic threshold: $f_{\rm crit}= 0.05
\sqrt{T/t_{200}}$.  The $\sqrt{T}$ time dependence was designed to
describe the diffusion of chaotic orbits as a random-walk through
phase space (Valluri \& Merritt 1998). This threshold isolated orbits
that were so chaotic that the orbital shape was strongly altered over
two successive time intervals.

In the set of calculations for Paper 1, the precise choice of $f_{\rm
crit}$ was not important, since nearly every orbit was stable and
therefore had a negligible $\Delta f$. However, adding a black hole
amplifies the level of chaos in a model, yields a wide spectrum of
$\Delta f$s, and requires a more careful study of chaotic thresholds.
In particular, histograms of chaotic subsamples preliminarily selected
with the previous criterion did not show a strong (i.e. $\sqrt T$)
diffusion signal. Concerned that the rising threshold would
underestimate the level of chaos in our models, we devised a new,
simpler technique to identify chaotic orbits.  In this paper, we
identify chaos as $\Delta f = \mid f_1 - f_2 \mid/ f_1 > f_{\rm
crit}$, where $f_{\rm crit}$ is a constant empirical threshold that
corresponds to the $95$th percentile of the $\Delta f$ distribution
for the most bound sample (at $E=-1.0$) {\it before} the black hole is
grown. For that sample, $\Delta f_{\rm crit}=0.1$.  Choosing the
threshold in this manner provides us with a comparative measure of the
black hole's effect on the phase space distribution. Additionally, the
threshold was high enough to ensure that the orbits we have identified
as chaotic did not arise from the much slower diffusion of orbits due
to potential noise or integration error. We tested the sensitivity of
our results by sliding the threshold from $0.5f_{\rm crit}$ to
$2.0f_{\rm crit}$, and found that the percentage of chaotic orbits was
unaffected in this range. In other words, chaotic orbits in this
sample typically had $\Delta f$ far larger than the threshold set by
any of these techniques.

\subsection{Planar orbit structure}

Figure 4 shows surfaces of section as a function of binding energy for
orbits confined to the $xy$ and $xz$ planes. As in Paper 1, orbits
were sampled to populate the surface of section evenly. In the inner
regions ($-1.0<E<-0.65$), the box and boxlet phase space is entirely
replaced by chaotic orbits.  In addition, there is an easily
discernible population of eccentric tubes that are also driven chaotic ---
on the surface of section, these orbits occupy the high velocity
boundary at a given initial position. While the tubes with larger
pericenters were unaffected by chaos, eccentric tubes all sample more of 
 the center of the potential, and are more likely to be driven chaotic. We note that the stable loops seen
in these tightly bound slices are not a result of motion inside the
black hole's Keplerian potential; even at $E=-1.0$, the mean
ellipsoidal radius $\langle q \rangle = 0.2 $ is far larger than the
range of influence of the black hole.

The outer regions of the model responded to the growth of the black
hole in a strikingly different manner. The number of orbits identified
as strongly chaotic fell dramatically compared to the inner regions,
and these chaotic orbits were randomly situated in box and boxlet
phase space. The 2:1 resonant boxlet (``banana'') is prominent and
stable in the $xz$ plane of this model, although it does not occur in
the original model without a black hole (see Miralda-Escude \& 
Schwarzschild 1989 or Lees \& Schwarzschild 1992 for 
more information on boxlet nomenclature).
In the $xy$ plane, the
resonant orbit islands are very weak, particularly in the $E=-0.40$
slice where chaotic and regular resonant orbits seem to be mixed
randomly throughout box phase space. The relative number of these
scattered ``boxlet'' orbits is consistent with the orbit
identification error of $1\%$.

A common interpretation for the lack of strong chaos in the weakly
bound orbits is that they have been integrated for far fewer dynamical
times than the highly bound orbits. For example, orbits in the
innermost slice (where $\langle q \rangle = 0.2$) were integrated for
$\approx 200 t_{\rm orb}$, while orbits in the outermost slice (with
$\langle q \rangle = 3.9$) had typically reached only $\approx 50
t_{\rm orb}$ before the hard limit set by the analysis routine.  Thus,
under this interpretation, there has simply been less time to observe
substantial stochastic diffusion in these outermost orbits, and a
longer integration might produce the same fraction of chaotic orbits
as is seen in the more tightly bound set.  To test this hypothesis, we
integrated a subset of the lesser bound box and boxlet orbits
($E=-0.40; \langle q \rangle = 1.3$) in the $xz$ plane for $\approx
625 $ orbital times (see Figure 5). After 200 orbital times, the
orbits were as dynamically evolved as the most tightly bound orbits,
but the fraction of box and boxlet phase space that had gone chaotic
only reached $7\%$ (versus $\approx 100\%$ of the box and boxlet phase
space at $E=-1.0$). As the integration time increased to well over two
Hubble times, the fraction of chaotic orbits also went up. However,
although the percentage of chaotic orbits in this subset increased
from $2\%$ to $16\%$ over the course of the experiment, even after a
scaled time of $29$ Gyr, many stable centrophilic orbits persist,
including a population of non-resonant boxes.  To test the extent of
numerical error over long integrations, we also integrated the same
subset of orbits (i.e. $xz$ plane, $E=-0.40$) from our initial
triaxial model without the black hole and in this experiment, only
$4\%$ of the orbits were considered strongly chaotic well after two
Hubble times.  So, the chaos present in these weakly bound orbits is a
real effect, not simply an artifact of the longer integration time.
Furthermore, the long integration time ensured that the remaining
stable box orbits were stable despite repeated passes through the
black hole's sphere of influence.

\subsection{Three-dimensional orbit structure}

To study the orbital evolution induced by a central black hole, it is
not sufficient to consider only two-dimensional surfaces of
section. Orbits in triaxial galaxies are not simply confined to the
major and minor planes; there are several important 3-D resonances
which serve to support a triaxial figure and which do not project down
onto an identifiable resonance in a principal plane. Furthermore, the
fact that 2-D orbits are confined to a plane dampens the onset of
chaos. For example, in a singular logarithmic triaxial potential, much
of the chaos present is in the 3-D orbits; this chaos is not reflected
in the orbit families moving in the principle planes
(e.g. Papaphilippou \& Laskar 1996).

With this in mind, we also investigated the three-dimensional orbital
content as traced by the particle distribution. As in Paper 1, we
sorted the final particle distribution according to binding energy and
binned the distribution into 9 slices of equal particle number. Figure
6 shows the percentage, by mass, of tube families versus chaotic orbits as a
function of binding energy and radius. Notice that the predominant
orbit families in this model are tubes, and the precipitous drop-off in
the number of chaotic orbits with decreasing binding energy, as seen
in the planar sample, is also reflected here.

Figure 7 presents the particle distribution on a frequency map of
$f_y/f_z$ versus $f_x/f_z$ after 50 orbital times. Since the initial
conditions were set by the actual N-body model, this frequency map is
not evenly sampled. Furthermore, since only the initial frequencies
are shown, this plot does not show orbital chaos. What is shown,
however, are the resonant regions that are populated by the particles
in this model.  In the most bound slice ($E=-1.0$), it is clear that
nearly the entire population is composed of tube orbits, since
practically all of the 40000 orbits in this slice lie along the
$(0,1,-1)$ or $(1,-1,0)$ resonance lines, marking the inner long-axis tubes 
and short-axis tubes, respectively.  In fact, the only
significant area that does not lie along these lines also contains
tube orbits; the clump near $(f_x/f_z, \space f_y/f_z) = (0.8,0.98)$,
contains orbits which project to long axis tubes in the xy-plane plane and low-order
boxlets in another. The majority of these tube orbits are well outside
the Keplerian potential of the black hole; these tubes are instead
dictated by the nearly spherical $stellar$ potential which extends out
past $\log q = -1$.  To underscore this point, we note that most of
these tubes in this energy slice clump near the $(1,1,1)$ resonance,
suggesting that the potential is not just axisymmetric here, but
nearly spherical (However we note that, {\it by itself}, a predominance
of 1:1:1 resonant orbits does not necessarily mandate a spherical
density distribution (Statler 1987; Hunter \& de Zeeuw 1992).)

As we move out in the model to the least tightly
bound particles, although the strong presence of tube orbits remains,
low order boxlets are discernible, as in the $(1,-2,1)$ resonance at
$E=-0.65$, and the $(2,0,-3)$ resonance at $E=-0.2$ and $-0.4$. These
final two panels strongly resemble the frequency maps for the same
energy slices in the model without a central black hole (see Paper
1). This, along with the lack of shape evolution beyond $\log q = 0$,
indicates that the effect of the black hole does not strongly alter
the outer regions of the model over galactic evolutionary timescales.

The general result that more loosely bound box orbits persist even
well after a Hubble time is not new. Both analytic estimates and early
numerical simulations indicated that while an unsoftened black hole
induces stochasticity in nearly all of the box and boxlet orbits, the
loss of triaxiality is confined to the center. For example, Gerhard \&
Binney (1985) modeled the disruption of planar box orbits as a series
of discrete scattering events by a $M_{\rm BH}= 0.02 M_{\rm core}$
black hole and found that, in a Hubble time, box orbits extending
far outside the core are unlikely to have experienced enough
pericentric passes to have been substantially scattered. In a fully
self-consistent N-body simulation of black hole growth in a
post-collapse triaxial potential, Norman, May \& van Albada (1985)
showed that a black hole causes box phase space to be replaced with
fully stochastic orbits on a timescale proportional to an orbit's
dynamical time. This was taken to be consistent with the black hole
scattering picture, with the caveat that the small number of particles
in the simulation could result in an additional numerical source of
scattering by two-body relaxation.

However, recent work has suggested that the trend toward stochasticity
is a more rapid and a more global phenomenon.  Merritt \& Quinlan
(1998) conducted self-consistent N-body experiments of black hole
growth in triaxial models, and determined that black hole masses
greater than $0.003M_{\rm gal}$ drive the galaxy toward axisymmetry,
and that global stochasticity -- where the axisymmetry extends well
outside the half-mass radius and evolves on the order of an orbital
period -- occurred at $M_{\rm BH}\approx 0.3 M_{\rm gal}$. Our more
subtle and local evolution is not necessarily at odds with these
results. The study conducted by Merritt \& Quinlan involved galaxies
which were more triaxial than ours, and which also had flat
($\gamma=0$) central density profiles. The different density structure
of the models results in different orbit populations in the two
studies -- the constant density cores of the Merritt \& Quinlan models
are likely populated by a much higher fraction of box orbits 
than our more cuspy models (Statler 1987; Hunter \& de Zeeuw 1992; 
Arnold \etal 1994). Furthermore, the growth
of a central black hole in a $\gamma=0$ model represents a more
significant perturbation to the central potential than in our
$\gamma=1$ models. Even with identical initial density profiles,
though, a direct comparison is complex. For example, using
Schwarzschild's (1979) method of reconstructing a $\gamma=0.5$ potential via
a judicious selection of orbit libraries, Valluri \& Merritt (1998)
found that slightly different initial minor axis flattenings give
different results for the relative fraction of tightly bound and less
bound chaotic orbits.  Moreover, the degeneracy between a given
physical shape and its possible orbital content makes any particular
solution for the orbital behavior non-unique.  It is apparent from our
study, though, that there exists at least one set of orbit families
that generate a triaxial figure and respond to an adiabatically
growing central potential in such a way as to preserve its global
shape over long timescales.

\subsection{Discussion}

It is clear that the onset of chaos in the most tightly bound orbit
families leads to a rapid change in the inner structure of the model
galaxies.  What is less clear is the specific agent driving the chaos
in the system --- is it scattering by the central black hole itself,
the effects of the steepened stellar density cusp, or something else
entirely?  One way to check if the stellar cusp is responsible for the
chaos is to remove the black hole and categorize the orbits that
result from the $\gamma \sim 2$ frozen stellar potential. Although
this is not a self-consistent galaxy model, it is a useful tool to
gauge the relative importance of the black hole to the steep stellar
cusp (and its spherical stellar potential).  We integrated a subset of
particles in the $E=-0.65$ energy slice in the purely stellar
potential for a total of $200 t_{\rm orb}$, and classified the orbits
in the manner described previously. In both the two- and
three-dimensional orbital analysis, we find that rapid chaos is
strongly suppressed when the black hole is removed. Therefore, it
seems clear from this experiment that the steep stellar cusp is not
the source of strong chaos in these models.  This result is
interesting in light of the work of Merritt \& Fridman (1996) which
links the rapid onset of global chaos to the presence of steep 
($\gamma \sim 2$) density cusps in galaxies. Earlier work 
by Schwarzschild (1993), however, demonstrated the secular stability 
of $\gamma \sim 2$ density cusps
in scale-free triaxial logarithmic potentials, more in line with our results.
These different conclusions indicate that
the connection between cusp slope, figure shape, and orbit
stochasticity may well be extremely sensitive to the various orbit
families initially present in the model.

While the presence of a black hole is necessary for driving chaos in
our models, the mechanics of this process are less clear. Under a
simple black hole scattering model, chaotic orbits should have more --
or closer -- pericentric passes within $r_{\rm BH}$ than stable ones,
but this is not the case in our sample.  In fact, the degree of
stochasticity depended on neither the number of pericentric passes
within $r_{\rm BH}$ nor on the minimum $r_{\rm peri}$. So, while the
black hole does induce strong chaos, it seems not to do so through a
simple scattering process. Another possibility is that chaos is driven
by the transition between the inner spherical Keplerian potential of
the black hole and the outer triaxial potential of the galaxy.  In
this picture, weakly bound box orbits in the outskirts of the model
may spend too little of their orbital period in the inflection region
to be significantly perturbed, and are thus not driven stochastic at
all. In our simulations, however, there was no strong correlation
between $\Delta f$ and the time spent within the inflection region,
leaving open the question of what drives chaos. In truth, however, the
graininess of the potential makes it difficult to disentangle the
subtleties of chaos from numerical noise in the potential, so that
analytic studies are better suited for this question.

In the outer regions, orbits stay regular even after repeated passages
near the potential center. It is possible that many of these orbits
are actually chaotic orbits that are ``sticky'' (Siopis \& Kandrup
2000), with a diffusion timescale that is much longer than a few
dynamical times. The course-grainedness of our potential seems to
argue against this explanation; a course-grained potential effectively
creates holes in the Arnold web (Arnold 1964) through which an
otherwise confined orbit may escape. In addition, we observed the same
effect in the planar orbit sample, and Arnold diffusion does not occur
in 2-D potentials (Merritt \& Fridman 1996). It seems likely, then,
that the regularity of these orbits is real.

Issues of orbital chaos and regularity aside, our galaxy model
maintains its original degree of triaxiality on a global scale despite
the presence of a massive central black hole. The orbit analysis
described in \S4 indicates that the model will remain relatively
stable for as long as a Hubble time. It is therefore interesting to 
compare this stable galaxy model with observations of real elliptical galaxies.

Initially, our scaling parameters were chosen to place the model on both the
global and the core fundamental planes. After the evolution
in structural and kinematic properties driven by the black hole, does
the model still obey these relations? In the case of the global
fundamental plane, the answer is certainly yes: the changes in the
velocity dispersion and radial density profiles occur only in the
central regions ( $ \log r < -1.5 $, or $ r $ \simlt 100  pc ), leaving the
global properties of the model unchanged.  For the core fundamental
plane, the increase in central cusp slope ( from $\gamma = 1$ to
$\gamma \sim 2$ ) represents a significant change, leaving the model
with a cusp slope which is perhaps a bit too steep for its scaled
luminosity of $M_B = -21.6$ (see \eg, Gebhardt \etal 1996). However,
the $\gamma-M_B$ relationship is quite steep and shows significant
scatter at intermediate luminosity (Gebhardt \etal 1996), so that this
discrepancy may not be significant -- ellipticals with $M_B \sim -21$
show cusp slopes of $\gamma = 1-2$. Scaling our model to {\it more} 
luminous ellipticals begins to present problems, however, since
adiabatic black hole growth models generically predict cusp slopes
steeper than that observed in luminous ellipticals (\eg Bahcall \&
Wolf 1976; Young 1980; Goodman \& Binney 1984; Sigurdsson \etal 1995;
Quinlan \etal 1995).

Turning to the issue of triaxiality in ellipticals, the triaxiality 
parameter of the
model at the half-mass radius is $T=(1-b^2)/(1-c^2)=0.5$, with a
flattening $c/a=0.8$. Unfortunately, a
problem arises in defining the best sample of ellipticals with which
to compare our model.  To select for black hole embedded ellipticals,
samples of radio-loud ellipticals might seem the best choice, but
luminous ellipticals typically have a flatter cusp slope than that
used our models. On the other hand, although a more general sample of
ellipticals may possess a wider range of cusp slopes, including
steeper cusps like that of our model, they may also harbor black
holes at a reduced rate when compared to AGN-selected samples.
Bearing these caveats in mind, we compare the structural properties of
our model to the triaxiality inferred for different samples of
elliptical galaxies. Using a combination of photometric and kinematic
data, Bak \& Statler (2000) show that the Davies \& Birkinshaw (1988)
sample of radio galaxies is characterized by a range of shapes biased
towards prolate figures but that -- as long as ellipticals are not
disk-like rotators -- triaxialities like that of our
model are common. In addition, the typical flattening of the galaxies
in that sample is $c/a \sim 0.7$, similar to, although somewhat
flatter than, our models.  Studies of larger samples of ellipticals
(not radio selected) also show characteristic flattening similar to
that of our model (e.g.  Ryden 1992; Tremblay \& Merritt 1995). While
a more detailed analysis of triaxiality in black hole embedded
ellipticals must await more complete data, at face value our model
does represent well the observed structural properties of elliptical
galaxies.

The connection between supermassive black holes and triaxiality has 
important consequences
for both secular and hierarchical galaxy evolution models.  For
example, it has been suggested that if a central black hole drives its
host galaxy toward axisymmetry globally and rapidly, one possible
difference between an intrinsically bright elliptical (thought to be
more triaxial) and a faint elliptical is that the stars in the faint
elliptical, with their shorter crossing times, have had more
interactions with the black hole and the galaxy is thus more
dynamically evolved (Valluri \& Merritt 1998). The black hole/bulge
mass relation can be explained in terms of galaxy evolution (Valluri
\& Merritt 1998) as well. In this scenario, spiral galaxies begin as
gas-rich disks with a small triaxial bulge. The triaxial potential
supports fueling of the central black hole through material falling
into it on box orbits (e.g. Norman \& Silk 1983) or by gas traveling
on intersecting orbits which drive dissipation and inflow.  The black
hole grows until a critical mass of $\simeq$ few $\%$ $M_{\rm gal}$,
which breaks triaxiality and strongly curtails the gas inflow.
Subsequent disk-disk merging can create a elliptical galaxy, and black
hole feeding ensues in this larger triaxial bulge until the critical
black hole mass is achieved. In both types of galaxies, the process is
the same: once the black hole mass fraction is large enough to disrupt
box orbits, gas inflow is sharply diminished.

The black hole in our model induced axisymmetry out to several hundred
parsecs, and resulted in a clearly observable change in the shape and
structure of the galaxy on these scales. Since the transformation did
not take place {\it globally}, it is tempting to say that the black
hole mass/bulge mass relation observed in the current galaxy
population is {\it not} simply an artifact of gas inflow in a more
triaxial-shaped progenitor population.  However, it is not immediately
clear how the more localized axisymmetry we observed would affect gas
inflow and subsequent black hole feeding. While it is true in our
globally triaxial model that gas inflow from outside the half mass
radius would never be entirely cut off, the behavior of the gas once
it hits the axisymmetric region requires detailed gas dynamical
simulations.  Nonetheless, it is clear that a central supermassive
black hole causes dramatic and long-lasting changes in the host galaxy
over scales well outside the region in which it dominates the
potential.

Finally, 21-cm observations of elliptical galaxies have revealed the
presence of extended neutral hydrogen disks and rings in many ellipticals (e.g.
Franx et al. 1994; van Gorkom \& Schiminovich 1997; Hibbard et
al. 2001).  Several authors have proposed that these structures are
precursors to ``disk rebuilding'' in ellipticals (Schweizer 1998; van
Gorkom 2001). If this scenario is correct, there must be an inward
migration of gas in these systems. Triaxiality offers a mechanism to
drive this inflow, but if black holes were to break triaxiality on
large scales it is difficult to see how migration and disk building
could occur. Our results alleviate these concerns; the fact that
triaxiality is maintained in all but the inner few hundred parsecs of
the galaxy would allow gas to move inwards on kiloparsec scales to
perhaps begin the process of disk formation. However, the modest
degree of triaxiality in the model suggests that the rebuilding
timescale may be long.\footnote{This is particularly problematic for the very
extended (\ie many $R_e$) rings seen in some ellipticals. In these
cases, not only is the dynamical timescale long, but the dynamics are
more likely driven by the structure of the dark matter halo than by the
luminous galaxy itself (\eg Franx, van Gorkom, \& de Zeeuw 1994).}
   
\section{Summary}

Using numerical simulations, we have studied the growth of central
supermassive black holes inside cuspy triaxial galaxies. Unlike previous
self-consistent modeling of black hole growth in triaxial ellipticals,
our calculations employ progenitor galaxy models which are both
triaxial and have central density cusps typical of observed
ellipticals (Holley-Bockelmann et al. 2000). Inside these progenitors,
we adiabatically grow a central black hole of mass $M_{\rm BH}=0.01 M_{\rm
gal}$.  As the black hole grows, it induces a central cusp ($\gamma
\sim 2$) in the stellar density profile and a strong roundening of the
central shape of the elliptical.  However, while the effects of the
black hole do extend beyond its nominal ``sphere of influence'', out
at an effective radius the galaxy figure remains largely unchanged.

To explore the change in the orbital structure induced by the black
hole, we use a combination of Fourier spectral classification and
axis-crossing pattern recognition to classify the orbits present in
the model. At the most tightly bound energies, the models are composed
entirely of loops, short and long axis tubes, and chaotic orbits. 
These chaotic orbits comprise
all of the box and boxlet phase space present in the original model,
and even a population of eccentric tubes. The outer regions
predominantly contain short axis tube orbits; however, in these outer regions,
there still exists a modest phase space of boxes and boxlets which
support the large-scale triaxiality of the system. Of these
centrophilic orbit families, there are a substantial fraction that do
{\it not} go strongly chaotic, even over thousands of dynamical times.
While the presence of noise in the potential expansion limits our
ability to detect mild chaos in the orbit populations, we believe that
the remaining boxes and boxlets in the outer regions are stable enough
to continue to support the global shape of the galaxy for a Hubble
time.

While massive black holes act as agents of chaos in the inner regions
of our models, they provoke
a more modest response in the outer regions than expected on the basis
of earlier studies (e.g. Norman et al. 1985; Merritt \& Quinlan
1998). However, those previous studies employed galaxy models which
had flat central density ``cores'' and were also significantly more
flattened than ours, a combination which yields a significantly higher
fraction of boxes in the original orbital families than found in our
model. These differences are likely the root cause of the more
dramatic evolution seen in previous calculations. Our
galaxy models indicate that even moderately triaxial ellipticals can
host central massive black holes, in agreement with observational
evidence which suggests both that black holes are ubiquitous
(Magorrian et al. 1998) and that triaxiality may be common (Bak \&
Statler 2000).

Our more realistic elliptical galaxy models make an excellent tool for the simulation
of several unsolved problems in elliptical galaxy formation and
evolution. such as the degree of disk rebuilding by large-scale gas
inflow, the effect of satellite infall on the structure of the galaxy,
the interaction between galaxies and their triaxial dark matter halos,
the persistence of central cusps, and the formation of nuclear disks.

\acknowledgements

This work is supported through a grant of computing time from the
National Partnership for Advanced Computational Infrastructure and the
San Diego Supercomputer Center, and by NASA through grant NAG5-7019. 
JCM is also supported by an NSF Career Grant and Research Corporation 
Cottrell Scholarship. We wish to thank Tim de Zeeuw for the 
careful refereeing of this paper. We thank Rebecca Stanek for 
help with data analysis.
 

\clearpage

\begin{figure}
\plotone{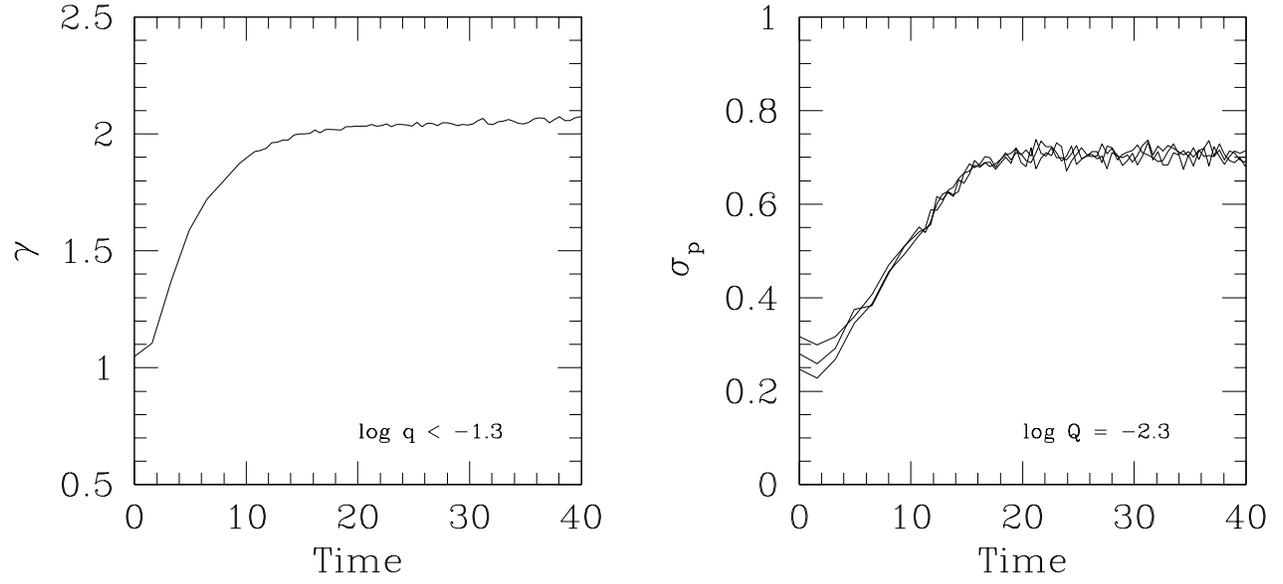}
\caption{Central properties of the model galaxy during and after black hole
growth. Left panel: Central density slope $\gamma$ of the model, measured at
ellipsoidal radius $\log q = -1.3$, as a function of time. Right panel: 
Central projected velocity dispersion, measured at projected ellipsoidal
radius $\log Q = -2.3$ as a function of time.}
\end{figure}

\begin{figure}
\plotone{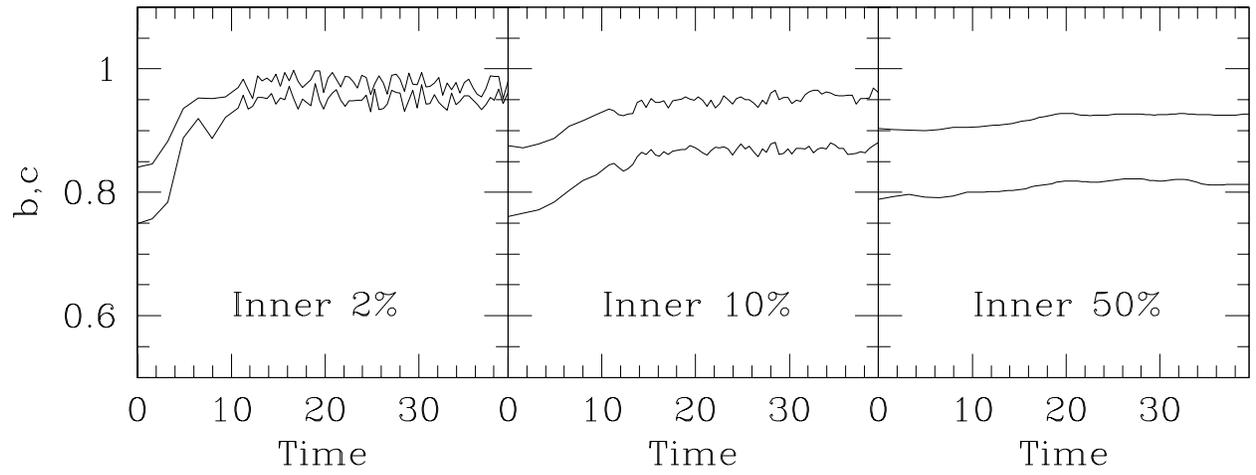}
\caption{The intermediate and minor axis lengths as a function of time 
for particle sets binned by mass in the model galaxy.  Axis lengths 
are calculated iteratively from the ellipsoidal density distribution 
using the moment of inertia tensor. 
(See Paper 1 for details.)}
\end{figure}

\begin{figure}
\plotone{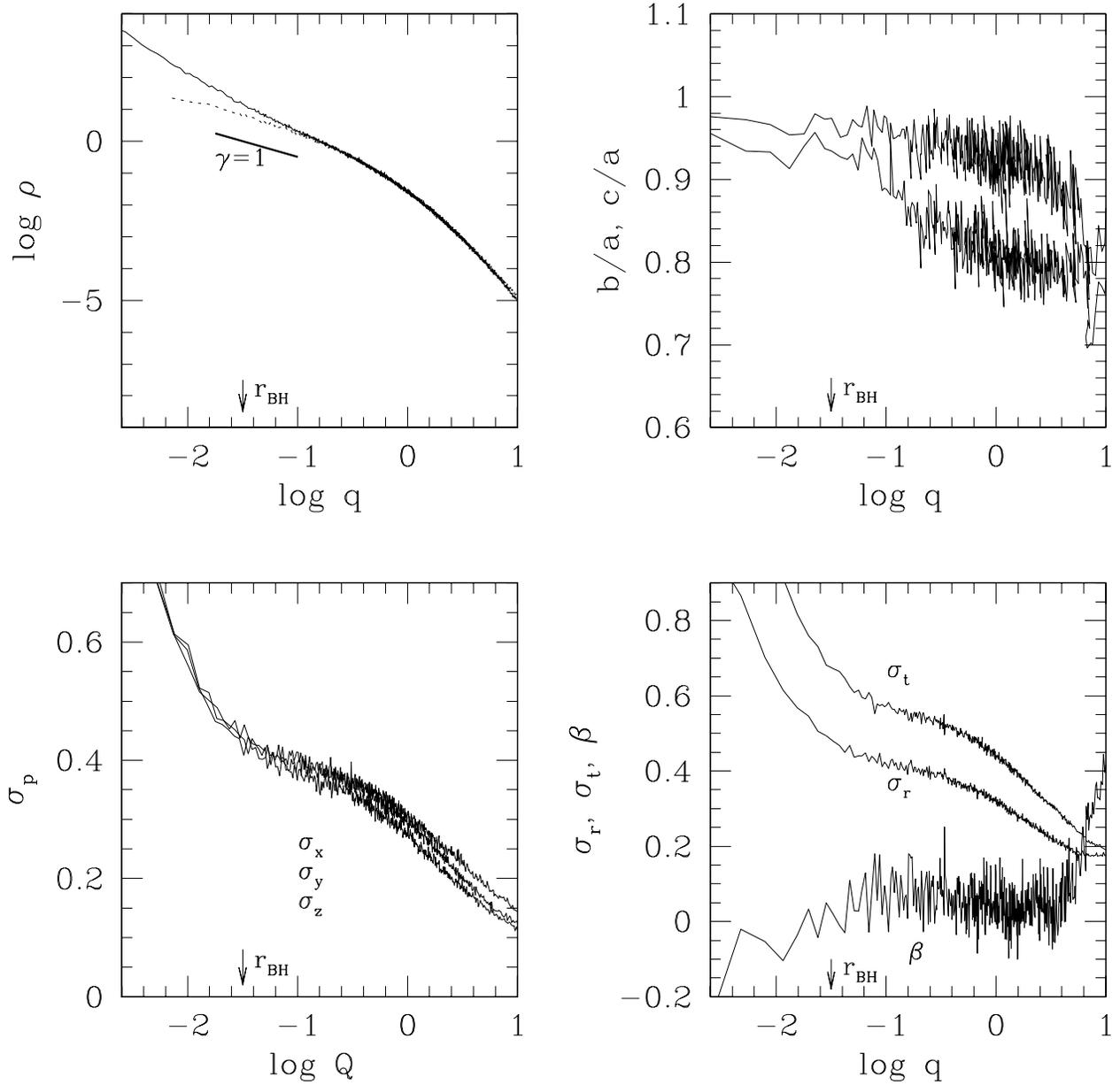}
\caption{The structural and kinematic properties of the model at $T=40$.
Upper left: density profile. Upper right: intermediate and minor axis lengths 
as a function of ellipsoidal radius. Lower left: projected velocity dispersion
along the fundamental axes, as a function of projected ellipsoidal radius.
Lower right: true radial and tangential velocity dispersion, and velocity
anisotropy parameter, as a function of ellipsoidal radius.}
\end{figure}

\begin{figure}
\plotone{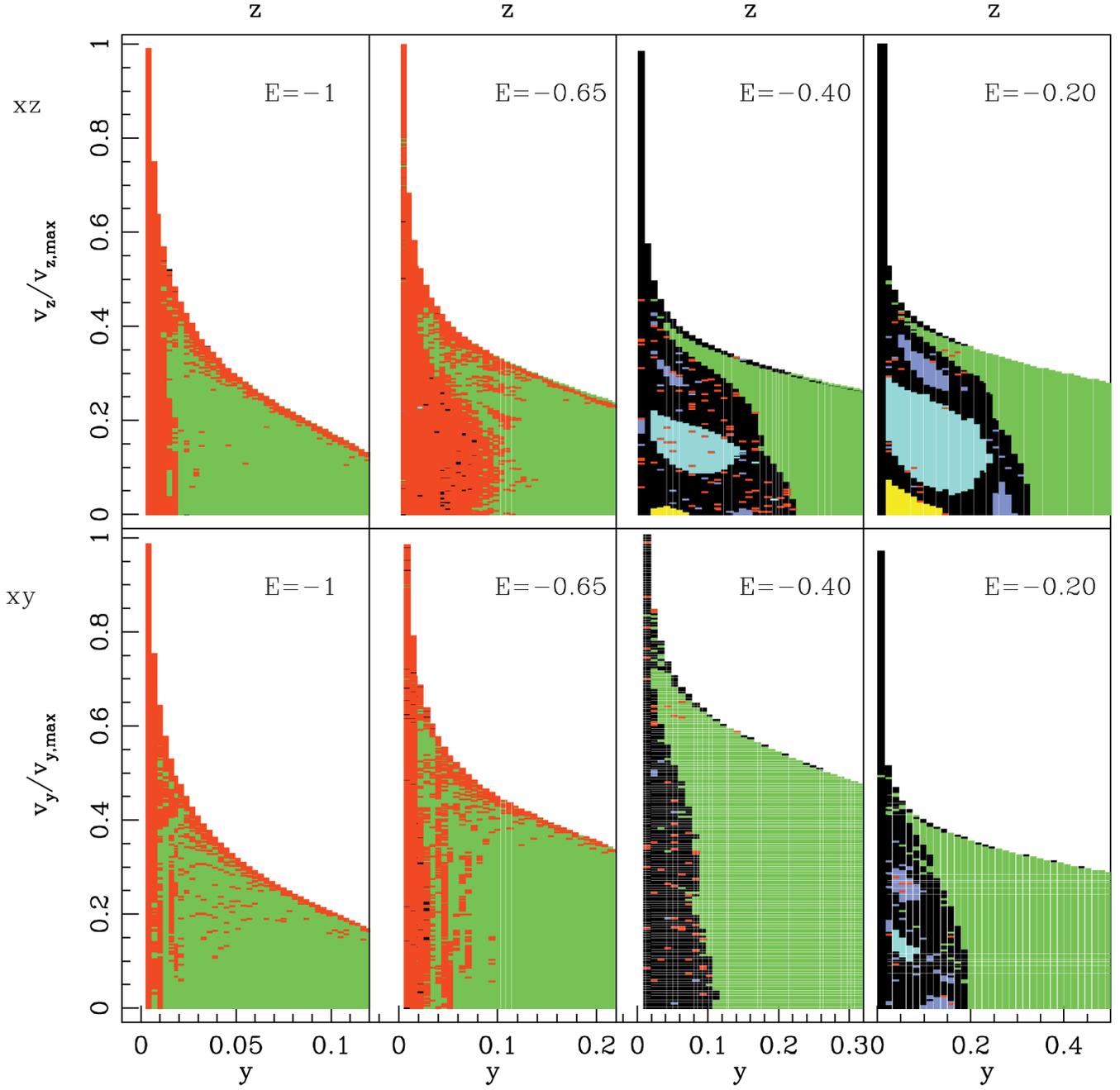}
\caption{Surfaces of section for the triaxial model at $T=40$,
plotted for orbital populations at different binding energies. Top:
Surfaces of section for orbits in the $xz$ plane. Bottom: Surfaces of
section for orbits in the $xy$ plane. Orbits are coded by color --
chaotic orbits: red; loops: green; bananas: yellow; fish: blue;
pretzels: aqua; higher resonance boxlets and pure boxes: black. This
plot was created by taking an average of all orbit types at a
particular position on the surface of section.}
\end{figure}

\begin{figure}
\plotone{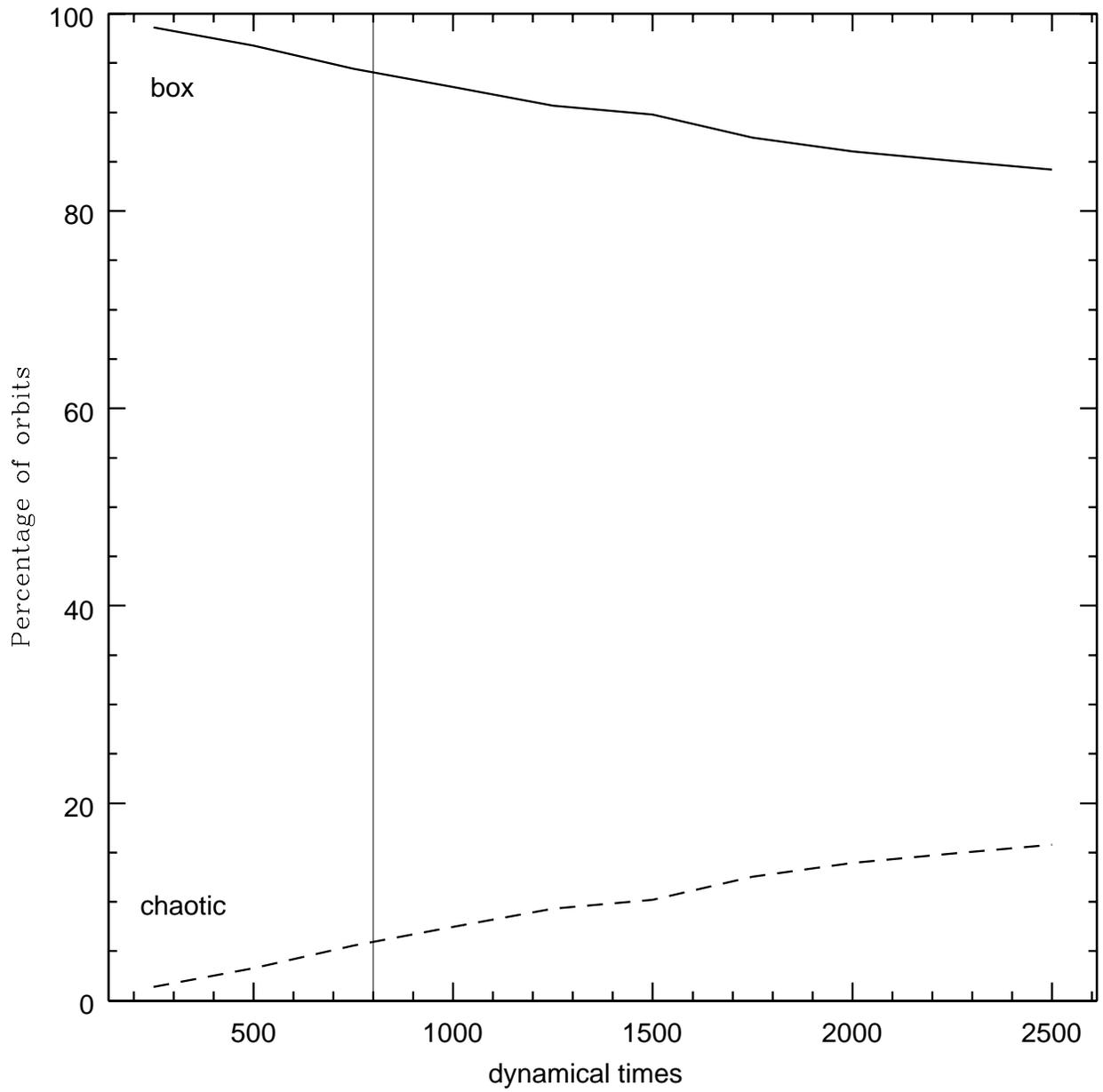}
\caption{The percentage of chaotic orbits as a function of dynamical time
for a subset of centrophilic planar $xz$ orbits at $E=-0.6$.}
\end{figure}

\begin{figure}
\plotone{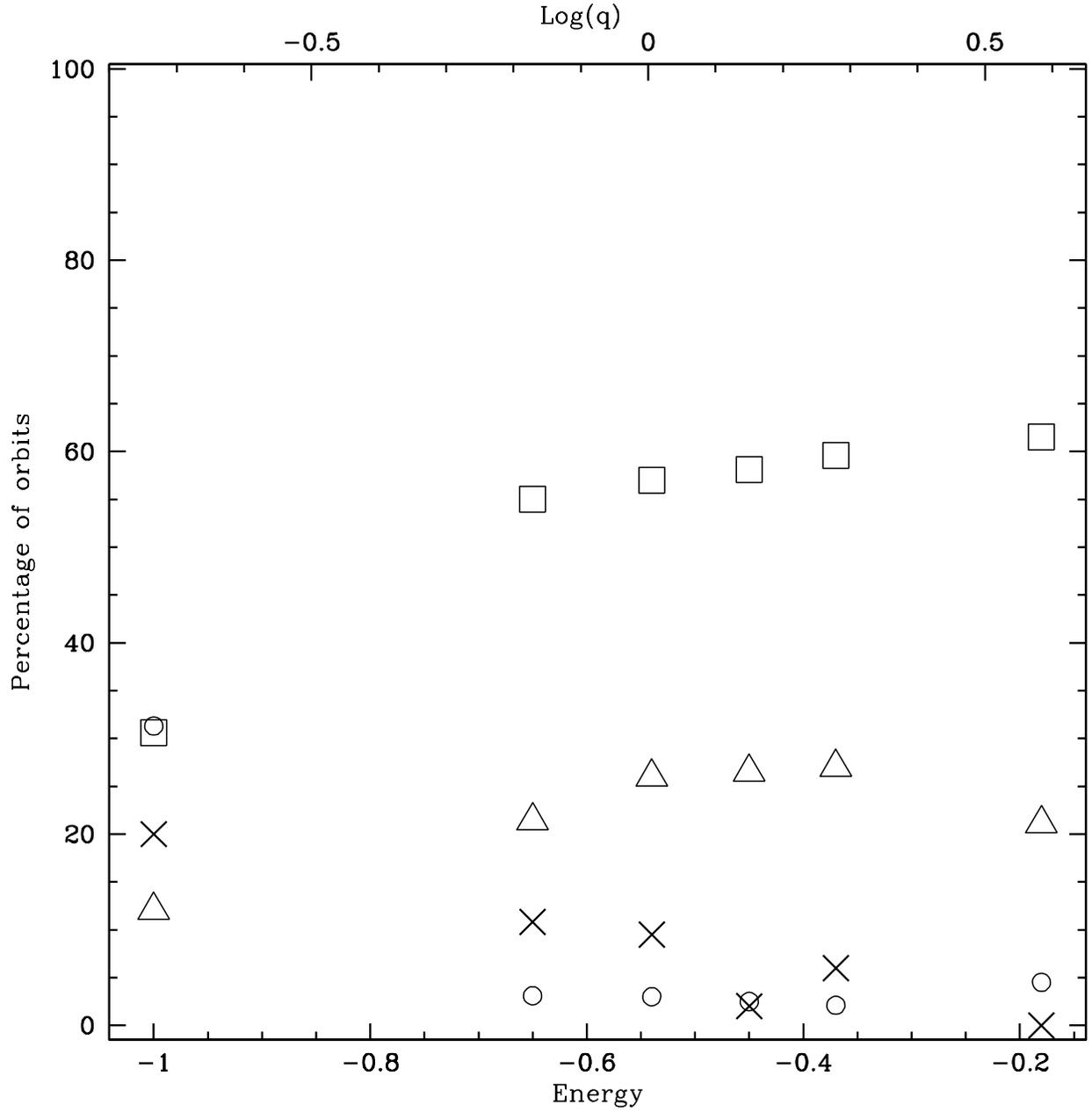}
\caption{Percentage of tubes vs. chaotic orbits in the 3-D sample as
a function of binding energy. Open symbols represent tubes in general; open squares represent short axis tubes, open triangles show long axis tubes, and open circles show the 1:1:1 resonance, or planar loop family. The crosses
represent strongly chaotic orbits.}
\end{figure}

\begin{figure}
\plotone{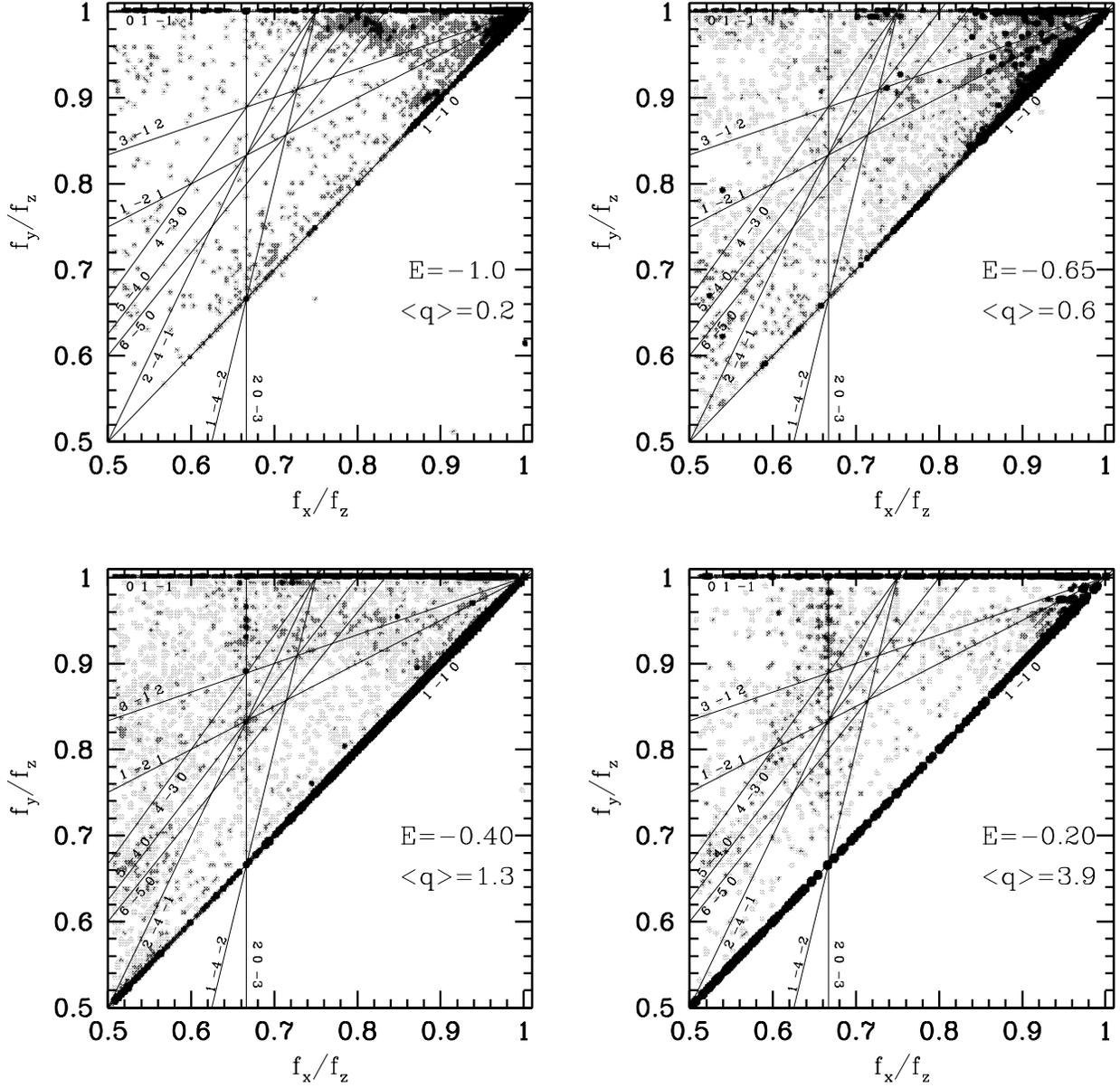}
\caption{Frequency map for the triaxial model at $T=40$,
plotted for orbital populations of different binding energies. The
greyscale represents the number of orbits at a given frequency ratio. 
The lightest grey is 1 orbit, while black is greater than 50 orbits.
The diagonal line corresponds to short axis tubes, and the horizontal
line at $f_y/f_z=1.0$ corresponds to long axis tubes.}
\end{figure}

\clearpage

\end{document}